\documentclass[12pt, border=55mm, preprintnumbers]{article}

\usepackage{amssymb}
\usepackage{amsmath}

\usepackage{graphicx}
\usepackage{color}
\usepackage{relsize}
\usepackage{lmodern}

\usepackage{float}

\def\bea{\begin{eqnarray}}
\def\eea{\end{eqnarray}}
\def\bean{\begin{eqnarray*}}
\def\eean{\end{eqnarray*}}

\begin{document}

{ \hspace{97mm} {\footnotesize{UCI-HEP-TR-2015-09}}}

\thispagestyle{empty}
\noindent\
\\
\\
\\
\begin{center}
\large \bf Minimal supersymmetric
standard model 
with \\ gauged baryon and lepton numbers\footnote{Plenary talk given at the International Conference on Massive Neutrinos, Singapore, February 9--13, 2015.}
\end{center}
\hfill
 \vspace*{1cm}
\noindent
\begin{center}
{\bf Bartosz Fornal}\\ \vspace{2mm}
{\emph{Department of Physics and Astronomy, \\ University of California, Irvine, CA 92697, USA}}
\vspace*{0.5cm}
\end{center}

\begin{abstract}
A simple extension of the minimal supersymmetric standard model in which baryon and lepton numbers are local gauge symmetries spontaneously broken at the supersymmetry scale is reported. This theory provides a natural explanation for proton stability. Despite violating $R$-parity, it contains a dark matter candidate carrying baryon number that can be searched for in direct detection experiments. The model accommodates a light active neutrino spectrum and predicts one heavy and two light sterile neutrinos. It also allows for lepton number violating processes testable at the Large Hadron Collider. 
\end{abstract}



\newpage 

\section{Introduction}
There have been several attempts in the 70's, 80's and 90's at promoting the accidental baryon and lepton number global symmetries of the standard model (SM) to local gauge symmetries (see Refs.~\cite{Pais:1973mi, Rajpoot:1987yg, Foot:1989ts, Carone:1995pu, Georgi:1996ei}). The first complete nonsupersymmetric model with gauged baryon ($B$) and lepton number ($L$) was constructed by P. Fileviez P\'{e}rez and M. B. Wise in Ref.~\cite{FileviezPerez:2010gw}. More realistic nonsupersymmetric models avoiding all current experimental constraints were introduced by M. Duerr, P. Fileviez P\'{e}rez and M. B. Wise in Ref.~\cite{Duerr:2013dza} and by P. Fileviez P\'{e}rez, S. Ohmer and H. H. Patel in Ref.~\cite{Perez:2014qfa}. The former model was further generalized and analyzed by M. Duerr and P. Fileviez P\'{e}rez in Ref.~\cite{Duerr:2014wra}. Its full supersymmetric version was constructed in Ref.~\cite{Arnold:2013qja} 
within the collaboration between
Pavel Fileviez P\'{e}rez, Sogee Spinner,
Jonathan M. Arnold and me, and is the subject of these proceedings.

The minimal supersymmetric standard model (MSSM) extended by right-handed (RH) neutrino superfields is one of the most promising candidates for physics beyond the standard model. It has a number of attractive features: it does not suffer from a large hierarchy problem, contains a dark matter (DM) candidate, and explains gauge coupling unification. However, despite all of those virtues, the MSSM also has its weaknesses. For instance, the discrete symmetry, $R$-parity, 
\bea
{R = (-1)^{3(B-L)+2s}} \ ,
\eea
needed to forbid proton decay at tree level and assure the stability of DM, is simply imposed by hand. Another uncomfortable aspect of the MSSM is connected to proton decay through nonrenormalizable operators. In particular, even after enforcing $R$-parity, there exist dangerous dimension-five operators triggering proton decay:
\bea
\frac{c_{\,1}}{\Lambda}\,\hat{Q}\hat{Q}\hat{Q}\hat{L} \ , \ \   \frac{c_{\,2}}{\Lambda}\,\hat{u}^c\hat{u}^c\hat{d}^c\hat{e}^c \ .
\eea
The constraint on the coefficient $c_{\,1}$ is model-dependent. However, assuming that all the couplings participating in the interaction are of order one sets the following upper bound on this coefficient:
\bea
|c_{\,1}| \  \lesssim  \ 10^{-25} \,\frac{m_{\rm soft} \,M_{\rm GUT}}{(1 \ {\rm TeV})^2} \ \approx \ 10^{-9}\ .
\eea
This number is small enough to be viewed as a problem. It turns out that gauging baryon and lepton number solves both of those issues.

\section{The model}
The theory is based on the SM gauge group extended by the additional baryon and lepton number gauge symmetries:
\bea
SU(3)_c \times SU(2)_L \times U(1)_Y \times U(1)_B \times U(1)_L \ .
\eea 
Without new particles, such a theory with just the MSSM particle spectrum is inconsistent, since it contains gauge anomalies. There are many ways to add new fields to make the theory anomaly-free. However, driven by the requirement of minimality, we choose to cancel the anomalies with the set of new fields given in Table 1, which introduces the minimum possible number of new degrees of freedom. 
\vspace{2mm}
{\renewcommand{\arraystretch}{1.4}
\begin{table}[h!]
\label{table:1}
   \begin{center}
   \begin{tabular}{| c || c | c | c | c | c |}
    \hline
        \rm{field}    & \raisebox{0ex}[0pt]{$ \!   {SU(3)} \! $} & \raisebox{0ex}[0pt]{$ \!   {SU(2)_L} \!$} & \raisebox{0ex}[0pt]{$\!   {U(1)_Y} \!  $} & \raisebox{0ex}[0pt]{$ \!   {U(1)_B} \! $} & \raisebox{0ex}[0pt]{$  \!  {U(1)_L} \! $}  \\ \hline\hline
$  {\hat{\nu}_{i R}}$ & $ {1}$ & $ {1}$ & $ {0}$ & $ {0}$ & $  {1} $ \\ \hline
           $  {\hat{\Psi}}$ & $ {1}$ & $ {2}$ & $ {-1/2}$ & $  {B_1}$ & $  {L_1}$  \\ \hline
           \ $  {\hat{\Psi}^c}$ & $ {1}$ & $ {2}$ & $ {1/2}$ & $  {B_2}$ & $  {L_2}$  \\ \hline
           $  {\hat{\eta}}$ & $ {1}$ & $ {1}$ & $ {-1}$ & $  {-B_2}$ & $  {-L_2}$  \\ \hline
           \ $  {\hat{\eta}^c}$ & $ {1}$ & $ {1}$ & $ {1}$ & $  {-B_1}$ & $  {-L_1}$  \\ \hline
           $  {\hat{X}}$ & $ {1}$ & $ {1}$ & $ {0}$ & $  {-B_2}$ & $  {-L_2}$  \\ \hline
           \ $  {\hat{X}^c}$ & $ {1}$ & $ {1}$ & $ {0}$ & $  {-B_1}$ & $  {-L_1}$  \\ \hline
    \end{tabular}
 \end{center}
\caption{New particle content of the model.}
\end{table} 

Apart from the  RH neutrino superfields we add six new leptobaryonic superfields, each containing a fermionic and scalar component. They all carry baryon and lepton quantum numbers satisfying the conditions:
\bea
B_1+B_2 = -3 \hspace{8mm} {\rm and} \hspace{9mm} L_1+L_2 = -3 \ ,
\eea
which are required by anomaly cancellation. The new superfields are vector-like with respect to the SM. This is why they can have vector-like masses and avoid the stringent experimental constraints on chiral matter. Now, since the $\hat{X}$ superfield is a singlet under the SM gauge group, its fermionic and scalar components are possible candidates for DM. The other fields should be unstable since they carry nonzero electric charge.

In order for those fields to actually obtain vector-like masses, we have to introduce new Higgs superfields into the theory. Because of the anomaly cancellation conditions, those new Higgs superfields, apart from being SM singlets, must carry baryon and lepton number 3 or -3:
\bea
\hat S_1   =  (1,1,0,3,3) \
\ \ \ {\rm{and}} \ \ \ \ \hat S_2   =  (1,1,0,-3,-3) \ .
\eea
The scalar components of those superfields develop vacuum expectation values (VEVs) below the $B$ and $L$ breaking scale. Because of their charges, baryon and lepton number can only be broken by a multiple of three units, so the theory predicts that:

\vspace{-5mm}

\bean
\bf{\emph{Proton is stable.} }
\eean
More generally, there are no interactions violating baryon or lepton number at the renormalizable level. The least suppressed nonrenormalizable operators generating $B$ and $L$ violation are dimension fourteen:
\bea
 { W_{14} } &=&   {\frac{1}{\Lambda^{10}} }
		  {\Big[}
			  {c_1 \hat{S}_1 (\hat{u}^c \hat{u}^c \hat{d}^c \hat{e}^c)^3  + c_2 \hat{S}_1 (\hat{u}^c \hat{d}^c \hat{d}^c \hat{\nu}^c)^3
		+ c_3 \hat{S}_2 (\hat{Q}\hat{Q}\hat{Q}\hat{L})^3}
		  {\Big]} \ .
\eea
The superpotential generating masses for the particles in our model is the following:
\bean
  {{\cal{W}}=Y_u \hat{Q} \hat{H}_u \hat{u}^c + Y_d \hat{Q} \hat{H}_d \hat{d}^c + Y_e \hat{L} \hat{H}_d \hat{e}^c
+  Y_\nu \hat{L} \hat{H}_u \hat{\nu}^c + \mu \hat{H}_u \hat{H}_d}
\eean

\vspace{-8mm}
\bean
  {+\
	Y_1 \hat{\Psi} \hat{H}_d \hat{\eta}^c
	+ Y_2 \hat{\Psi} \hat{H}_u \hat{X}^c
	+ Y_3 \hat{\Psi}^c \hat{H}_u \hat{\eta}
	+ Y_4 \hat{\Psi}^c \hat{H}_d \hat{X}}
\eean

\vspace{-8mm}
\bea
	\   {\hspace{-4mm}+\  \lambda_1 \hat{\Psi}  \hat{\Psi}^c \hat{S}_{1}
	+ \lambda_2 \hat{\eta}  \hat{\eta}^c \hat{{S}}_{2}
	+ \lambda_3 \hat{X} \hat{X}^c \hat{{S}}_{2}
	+ \mu_{BL} \hat{S}_{1} \hat{{S}}_{2}} \ .
\eea
The first line consists of the usual Yukawa and Higgs terms responsible for the MSSM masses. The second line contains new Yukawa interactions between the leptobaryons. The third line provides vector-like masses for the leptobaryons and the new Higgses. There are no terms mixing the new sector with the MSSM. The new fields couple to the MSSM only through the gauge sector. Since there are no operators violating baryon or lepton number, $R$-parity does not have to be imposed to avoid proton decay. 

It turns out that both before and after $B$ and $L$ breaking, the Lagrangian has a discrete $Z_2$ symmetry -- it is invariant with respect to the following transformation of the new fields: 
\bea
\hspace{-5mm}\hspace{-5mm}&&{\hat\Psi \!\to\! - \hat\Psi, \hspace{3mm} \hat\Psi^c \!\to\! - \hat\Psi^c,} \hspace{3mm}
  {\,\hat\eta \!\to\! - \hat\eta, \hspace{3mm} \hat\eta^c \!\to\! - \hat\eta^c,} \hspace{3mm} {\hat{X} \!\to\! - \hat{X}, \hspace{3mm} \hat{X}^c \to - \hat{X}^c}  ,
\eea
under which all leptobaryons are odd and all MSSM fields are even. 
The consequence of this accidental $Z_2$ symmetry is that the lightest leptobaryon is stable, since it cannot decay to MSSM particles. If the lightest leptobaryon is also electrically neutral, like the $\hat{X}$ superfield, its fermionic and scalar components are possible DM candidates. In our analysis, we assumed that the fermionic component of $\hat{X}$, the $\tilde{X}$, is the DM. It interacts with the MSSM only through the gauge bosons and its mass, given by the VEV of the scalar component of $\hat{S}_2$, is set at the baryon and lepton number breaking scale. 

\section{Symmetry breaking}

The scalar potential relevant for $B$ and $L$ breaking is:
\bea
\hspace{-8mm}{V} &=&{( M_{1}^2 \!+\! |\mu_{BL}|^2 ) \,|S_{1}|^2 + (M_{2}^2 \!+\! |\mu_{BL}|^2 ) \,|{S}_{2}|^2   + \frac{9}{2} g_B^2 \,(|S_{1}|^2 \!-\!  |{S}_{2}|^2 )^2}\nonumber \\
&& {+  \   \frac{1}{2} g_L^2 (3 |S_{1}|^2 \!-\!  3 |{S}_{2}|^2  )^2  -  \left( b_{BL} S_{1} {S}_{2} \!+\! {\rm{h.c.}}\right)} \ .
\eea
After the scalar components of the new Higgs superfields $\hat{S}_1$ and $\hat{S}_2$ get their VEVs, $\langle S_1 \rangle \equiv v_1$ and $ \langle S_2 \rangle \equiv v_2$, the resulting mass matrix for the new $Z'$ gauge bosons has a zero determinant.
This signalizes the existence of a new massless gauge boson and is ruled out by experiment. It is simple to understand why this happens -- after $B$ and $L$ breaking there is still a residual $U(1)_{B-L}$ gauge symmetry,
\bea
U(1)_B \times U(1)_L \rightarrow U(1)_{B-L} \ ,
\eea
with a $Z'$ gauge boson that does not get a mass through the Higgs superfields.
A natural solution to this problem is to introduce a new superfield whose scalar component does get a VEV which breaks $B\!-\!L$. This is clearly an option, however, in order to keep the model as minimal as possible, one can use a field already existing in the theory. A perfect candidate is the  RH sneutrino, since it is a SM singlet and carries nonzero lepton charge, so it obviously breaks $B\!-\!L$. 
A nonzero VEV for the RH sneutrino,
\bea
\langle \tilde{\nu}_R \rangle \equiv v_R \ne 0 \ ,
\eea
produces additional terms in the scalar potential, 
\bea
\hspace{-22mm}{V} &\!\!=\!\!& {( M_{1}^2 \!+\! |\mu_{BL}|^2 ) \,|S_{1}|^2 + (M_{2}^2 \!+\! |\mu_{BL}|^2 ) \,|{S}_{2}|^2   + \frac{9}{2} g_B^2 \,(|S_{1}|^2 \!-\!  |{S}_{2}|^2 )^2}\nonumber \\
&& {+  \  M_{\tilde \nu^c}^2 | \tilde \nu^c |^2+ \frac{1}{2} g_L^2 (3 |S_{1}|^2 \!-\!  3 |{S}_{2}|^2 \!-\! | \tilde \nu^c |^2 )^2  -  \left( b_{BL} S_{1} {S}_{2} \!+\! {\rm{h.c.}}\right)} \ ,
\eea
which now gives a mass matrix for the $Z'$ gauge bosons of the form: \bea
\mathcal{M}_{Z'}^2 =
	9\begin{pmatrix}
		g_B^2 (v_1^2+v_2^2)
		&
		g_B g_L (v_1^2+v_2^2)
		\\
		g_B g_L (v_1^2+v_2^2)
		&
		g_L^2 (v_1^2+v_2^2) + \tfrac{1}{9}g_L^2 v_R^2
	\end{pmatrix} ,
\eea
with both eigenvalues positive.

The breaking of $B\!-\!L$ implies that $R$-parity, which is a discrete subgroup of $U(1)_{B-L}$, is also spontaneously broken. However, proton is still stable since the VEV of the RH sneutrino breaks only lepton number. Analyzing the symmetry breaking conditions for the scalar potential reveals that the VEVs: $v_1$, $v_2$ and $v_R$ have to be of the same order as the supersymmetry breaking soft terms, simply because they enter the potential on equal footing. Another way to phrase this is that $U(1)_B$, $U(1)_L$ and $R$-parity have to be broken at the SUSY scale, which we set around a TeV. The violation of $R$-parity at the TeV scale opens the door to many interesting and exotic signatures at the LHC, including decays of the LSPs. On top of that, since our model has a DM candidate not affected by $R$-parity violation, one expects an additional presence of missing energy signatures.

\section{Neutrino sector}
After $B$ and $L$ breaking the model reduces to an effective $B\!-\!L$ gauged model. It turns out that the breaking of the remnant $U(1)_{B-L}$ gauge symmetry by the VEV of the  RH sneutrino has very interesting consequences for neutrino physics. Those were already studied in the case of a pure $B\!-\!L$ gauged extension of the MSSM in Refs. \cite{Barger:2008wn, Barger:2010iv, Ghosh:2010hy, FileviezPerez:2012mj, Marshall:2014cwa}. The predictions for the neutrino sector are very similar here and we discuss them following the analysis done by V.~Barger, P.~Fileviez P\'{e}rez and S.~Spinner in Ref. \cite{Barger:2010iv}.

First of all, the existence of  RH neutrinos in our model allows for standard Dirac mass terms through Yukawa couplings. In addition, $R$-parity violation triggers mixing between the neutrinos and neutralinos, since after $B\!-\!L$ breaking they have common mass terms. This generates Majorana masses for the left-handed (LH) neutrinos. 

One can show that only one generation of  RH sneutrinos can attain a significant VEV. Since this VEV is at the SUSY breaking scale, we expect it to be around a TeV. The other  RH sneutrino VEVs have to be very small and we ignore them: 
\bea
 {\langle \tilde{\nu}_{R1} \rangle \equiv v_R \sim {\rm TeV} \ , \ \ \langle \tilde{\nu}_{R2}\rangle =  \langle \tilde{\nu}_{R3}\rangle = 0} \ .
\eea
The mixing between neutrinos and neutralinos comes from the gauge sector term  $g_{BL} v_R (\nu_R^c \tilde{B}')$   and the superpotential term $ Y_\nu v_R (l^T i \sigma_2 \tilde{H}_u)$. 
In principle, the situation is a little more complicated, since the LH sneutrinos can also develop VEVs. These VEVs, however, have to be very small, since they produce mixing between the LH neutrinos and neutralinos, which directly affects the active neutrino masses. 
This mixing arises from the terms:
$ g_{BL} v_L (\nu_L \tilde{B}')$,  $g_{1} v_L (\nu_L \tilde{B})$,  $g_{2} v_L (\nu_L \tilde{W}^0)$ and $Y_\nu v_L (\nu_R^c \tilde{H}_u)$.
Consistency with experiment requires:
\bea
 {\langle \tilde{\nu}_{Li} \rangle \equiv v_{Li} \lesssim {\rm MeV} } \ .
\eea

The contributions to the neutrino masses can be described by the $11\times11$ neutrino-neutralino mass matrix. In the basis 
$\left(\nu_{iL}, \nu^c_{jR}, \tilde{B}', \tilde{B}^0, \tilde{W}^0, \tilde{H}^0_d, \tilde{H}^0_u\right) $, i.e., the three active LH neutrinos, three sterile RH neutrinos and the five heavy gauginos and Higgsinos, this matrix takes the form:\\
\bea
 { \hspace{-0mm}\left(
\begin{smallmatrix}
    {\huge{0}}_{3\times3} & \frac{v_u}{\sqrt2}Y_\nu^{j1} &\frac{v_u}{\sqrt2}Y_\nu^{j2}& \frac{v_u}{\sqrt2}Y_\nu^{j3}& -\frac{1}{2}g_{BL} v_{Li} & -\frac{1}{2}g_{1} v_{Li} & \frac{1}{2}g_{2} v_{Li} & 0 & \frac{ v_{R}}{\sqrt2}Y_\nu^{i3} \\
     \frac{v_u}{\sqrt2}Y_\nu^{1j}& 0&0&0 & 0  & 0 & 0 & 0 & \frac{v_{Lj}}{\sqrt2}Y_\nu^{1j} \\
    \frac{v_u}{\sqrt2}Y_\nu^{2j} & 0&0&0 & 0  & 0 & 0 & 0 & \frac{ v_{Lj}}{\sqrt2}Y_\nu^{2j} \\
    \frac{v_u}{\sqrt2}Y_\nu^{3j}& 0&0&0 & \frac{1}{2}g_{BL} v_R & 0 & 0 & 0 & \frac{v_{Lj}}{\sqrt2}Y_\nu^{3j} \\
    -\frac{1}{2}g_{BL} v_{Li} & 0 &0 & \frac{1}{2}g_{BL} v_R   & M_{BL} & 0 & 0 & 0 & 0 \\
    -\frac{1}{2}g_{1} v_{Li} &0&0&0 & 0 & M_1 & 0 & -\frac{1}{2} g_1 v_d & \frac{1}{2} g_1 v_u \\
    \frac{1}{2}g_{2} v_{Li} & 0&0&0 & 0 & 0 & M_2 & \frac{1}{2} g_2 v_d & -\frac{1}{2} g_2 v_u \\
    0 & 0&0&0 &0& -\frac{1}{2} g_1 v_d & \frac{1}{2} g_2 v_d & 0 & -\mu \\
    \frac{v_{R}}{\sqrt2}Y_\nu^{3i} & \frac{v_{Lj}}{\sqrt2}Y_\nu^{j1} & \frac{v_{Lj}}{\sqrt2}Y_\nu^{j2}  & \frac{v_{Lj}}{\sqrt2}Y_\nu^{j3} & 0 & \frac{1}{2} g_1 v_u & -\frac{1}{2} g_2 v_u & -\mu & 0 \\
  \end{smallmatrix}
\right)}. \ \ \ \ \\ \nonumber
\eea
The $6\times6$ subblock involving the LH and  RH neutrinos contains just Dirac mass terms. Since only one generation of  RH sneutrinos develops a TeV-scale VEV, this implies that only one of the  RH neutrinos obtains a TeV mass through the seesaw mechanism. Thus, effectively, there is a $5\times5$ subblock for the light states (three active and two sterile neutrinos) on the order of $0.1 \rm{ \ eV}$, and a $6\times6$ TeV-scale subblock involving the neutralinos and the heavy  RH neutrino. 

The off-diagonal blocks take part in generating Majorana masses for the active neutrinos through an effective type I seesaw mechanism. Experimental limits on the active neutrino spectrum constrain those two subblocks to be at the MeV scale at most. All those constraints can be translated into upper bounds on the Yukawa couplings and the aforementioned VEV of the LH sneutrino: 
\bea
Y_\nu^{ij} \lesssim 10^{-12} , \ \ \ Y_\nu^{i3} \lesssim 10^{-6}, \ \ \ v_{Li} \lesssim 1 \ {\rm MeV}  \ .
\eea
The hierarchy of scales in the full $11\times11$ mass matrix:
\bea
 {\mathcal{M} = \left(
  \begin{array}{cc}
    (m)_{5\times5} & (m_D)_{5\times6} \\
    (m_{D})^{T}_{6\times 5} & (M_{\rm heavy})_{6\times6} \\
  \end{array}
\right) \sim \left(
      \begin{array}{cc}
        \lesssim 0.1 {\rm \ eV} & \lesssim {\rm MeV}  \\
        \lesssim {\rm MeV}  & \sim {\rm TeV}  \\
      \end{array}
    \right)}
\eea
enables us to diagonalize the $5\times5$ light neutrino mass matrix perturbatively, using the standard seesaw formula:
\bea
  {\mathcal{M}_{\rm light} = m - m_D (M_{\rm heavy})^{-1} m_D^T } \ .
\eea
Keeping only terms with a maximum product of two small flavorful parameters:
\bea
 {\frac{v_{Li}}{m_{\rm soft}} , \, \frac{Y_\nu^{i3}v_u}{m_{\rm soft}} \lesssim 10^{-3}} \ ,
\eea
yields the following mass matrix for the five light neutrino states:
\bea
\hspace{-1mm}  {\mathcal{M}_{\rm light} =\! \left(
  \begin{matrix}
    A \,v_{Li} v_{Lj} + B \,v_u\left(Y_\nu^{i3} v_{Lj} + Y_\nu^{j3} v_{Li}\right)  + C \,v_u^2 Y_\nu^{i3} Y_\nu^{j3}& \ \ \frac{1}{\sqrt2}Y_\nu^{i\beta} v_u \\
    \frac{1}{\sqrt2}Y_\nu^{\alpha i} v_u & \ \ 0_{2\times2} \\
  \end{matrix}
\right)}  .
\eea
The sterile neutrinos have no Majorana masses in this approximation. In the active neutrino $3\times3$ subblock the parameters $A$, $B$, $C$ are on the order of $ 1/m_{\rm soft}$ which is essentially 1/TeV. Surprisingly, the determinant of this $3\times3$ matrix is zero due to its flavor structure. This indicates that one of the active neutrinos has a vanishing Majorana mass.
Therefore, after partial diagonalization, we can write the $5\times5$ light neutrino mass matrix as:
\bea
{\renewcommand{\arraystretch}{1} {\mathcal{M}_{\rm light}^D = \left(
                            \begin{array}{ccc|cc}
                              m_1 & 0 & 0 &  &  \\
                              0 & m_2 & 0 &  &  \\
                              0 & 0 & 0 & \multicolumn{2}{c}{\smash{\raisebox{1.1\normalbaselineskip}{\mbox{$\frac{v_u}{\sqrt2}U^{ji} Y_\nu^{i\beta} $}}}}   \\ \hline
                               &  &  &  \ \ 0 &  \ \ 0 \\
                                \multicolumn{3}{c}{\smash{\raisebox{0.6\normalbaselineskip}{\mbox{$\frac{v_u}{\sqrt2}U^{ij}Y_\nu^{\alpha i}$}}}}    &\ \  0 &\ \  0
                            \end{array}
                          \right)}.}
\eea
The $U_{ij}$ are the elements of the $3\times3$ unitary matrix diagonalizing the active neutrino subblock. Therefore, without having to assume unnatural cancellations:
\bean
 \emph{The model implies the existence of two light sterile neutrinos.} 
 \eean
Depending on the magnitude of $m_1$, $m_2$ and the Yukawa couplings there are three possibilities:

\begin{itemize}
\item[(a)]
If the Majorana contributions are negligible, neutrinos obtain their masses only through Dirac terms. This case is generic for many models. However, for this particular model it is in tension with oscillation experiments since the active-sterile neutrino mixing is maximal and not all mass differences between the active and sterile neutrinos can be zero  due to the different number of light LH and RH fields. 
\item[(b)]
The second possibility involves comparable Dirac and Majorana masses. Most of those cases are ruled out by oscillation experiments since they lead to large active-sterile mixing with significant left-right mass differences. There are some special cases, for example:
\bea
 {\mathcal{M}_{\rm light}^D = \left(
                            \begin{smallmatrix}
                              m_1 & 0 & 0 & 0 &  0\\
                              0 & m_2 & 0 & 0 &  0  \\
                              0 & 0 & 0 & \frac{v_u}{\sqrt2}U^{3i} Y_\nu^{i1}  &  0  \\ 
                              0 &  0 &  \frac{v_u}{\sqrt2}U^{i3}Y_\nu^{1 i} & 0 & 0 \\
                               0 &  0 &  0   & 0 & 0
                            \end{smallmatrix}
                          \right)}  ,
                          \eea
which avoid a large mixing, but they require an unnatural hierarchy between the Yukawas.
\item[(c)]
The final possibility is when there exist only Majorana mass terms, i.e., the Yukawas are negligible. In this case our model has the most interesting consequences. One does not have to worry about constraints from oscillation experiments, since there is no active-sterile mixing. In addition, this situation is interesting from a collider point of view -- the Majorana masses are generated as a result of $R$-parity violation, which leads to lepton number violating signatures. Although this possibility requires all Yukawas to be quite small, there might exist an underlying, not yet discovered symmetry responsible for this. We therefore concentrate on this particular case below.
\end{itemize}

Since our model predicts one massless active neutrino, one can use the experimental values of solar and atmospheric neutrino mass splittings to write down the exact active neutrino spectrum, given the choice of hierarchy:
\begin{itemize} 
\item[$\bullet$] Normal hierarchy:
\bean
 \hspace{-7mm}{m_1}  = {0} \ , \ \ 
 {m_2}   = {\sqrt{\Delta m_{\rm sol}^2}\approx 9 \ {\rm meV}}  , \ \ 
 {m_3}  = {\sqrt{\Delta m_{\rm atm}^2}\approx 50 \ {\rm meV}} .
\eean
\item[$\bullet$] Inverted hierarchy:
\bean
 \hspace{-7mm} {m_3}  =  {0} \ , \ \ 
 {m_1} = {\sqrt{\Delta m_{\rm atm}^2}\approx 50 \ {\rm meV}}  , \ \ 
 {m_2}   = {\sqrt{\Delta m_{\rm atm}^2 + \Delta m_{\rm sol}^2}\approx 50 \ {\rm meV}} .
\eean
\end{itemize}
Regarding the two light sterile neutrinos, without sizable Dirac terms the model predicts that they have extremely small masses. As explained above, the spectrum contains also a TeV-scale sterile neutrino:
\bea
 {m_4, m_5}  \ll {{\rm max}\{m_1, m_2, m_3\}} \ , \ \ \ 
 {m_6} \sim  {{\rm TeV}} \ .
\eea
Finally, although the model does not make concrete predictions for the active neutrino mixing, one can express the values of the mixing angles $\theta_{ij}$ in terms of the LH sneutrino VEVs and the Yukawa couplings: 
\bea
\theta_{ij} \  \longleftrightarrow \ {v_{Li} , \ Y_\nu^{i3}} \ .
\eea
There exists a concrete set of values for those six parameters which reproduces the experimental results for the mixing angles.

\section{Leptobaryonic dark matter}

As emphasized earlier, the DM candidate in the model is the fermionic component of the  $\hat{X}$ superfield. Because of its quantum numbers, it couples to the MSSM only through the gauge bosons (new $Z'$s). In our analysis we assumed that it interacts predominantly with the leptophobic gauge boson $Z_B$.  There are two main constraints which have to be fulfilled. 

If one requires thermal production of DM, the $\tilde{X}$ particle must have an appropriate annihilation cross section to satisfy the relic density bounds. The process responsible for this is  shown in Fig.~\ref{fig:1}. 
The other constraint comes from direct detection experiments, which look for DM scattering off nuclei. This process is described by a similar diagram (see Fig.~\ref{fig:2}).

\vspace{2mm}

  \begin{minipage}{\linewidth}
      \centering
      \hspace{-10mm}
      \begin{minipage}{0.45\linewidth}
          \begin{figure}[H]
              \includegraphics[trim=0.0cm 0cm 0cm 0cm, clip=true, width=\linewidth]{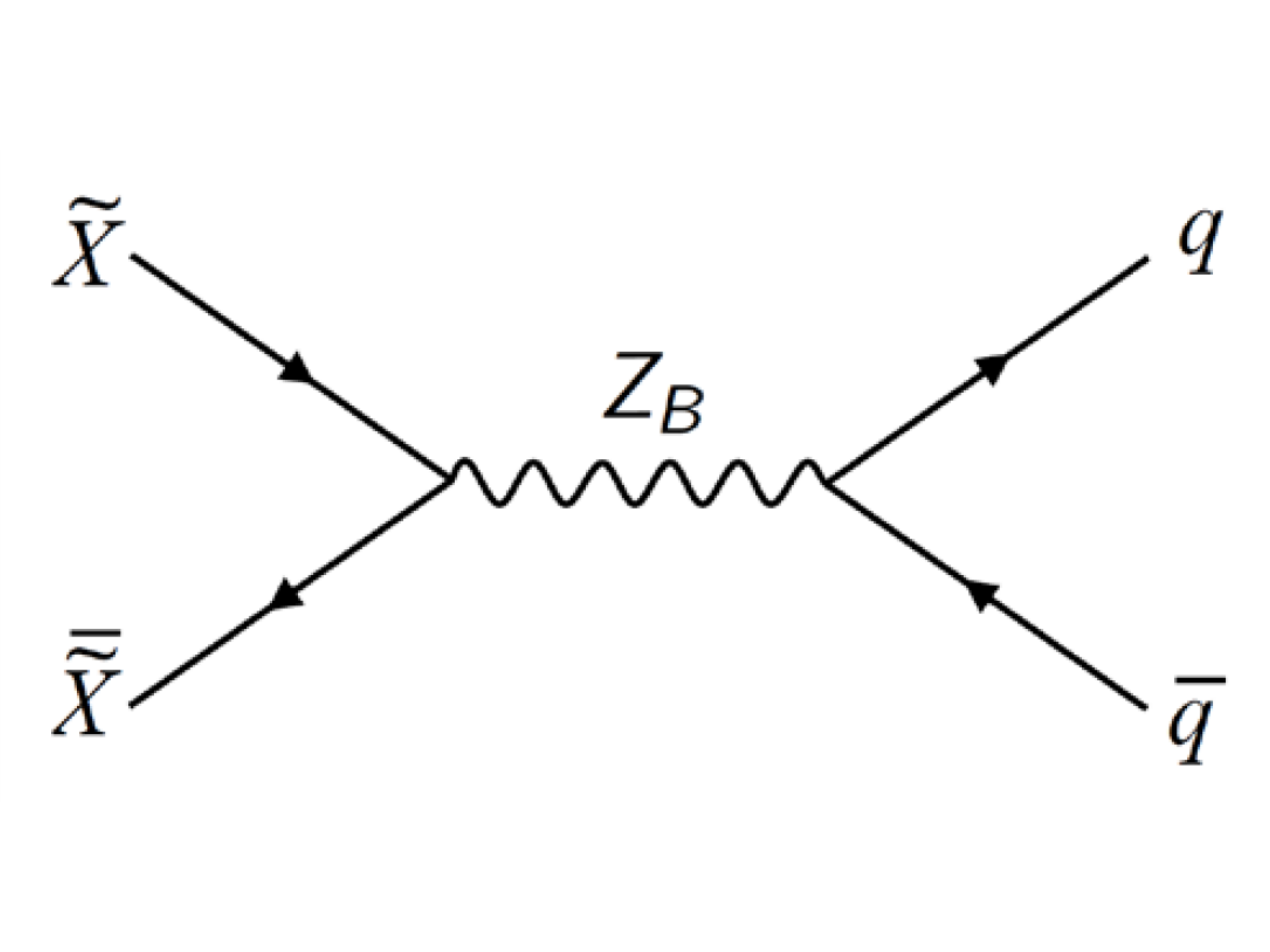}
\caption{Dark matter annihilation diagram $ {\bar{\tilde{X}} \tilde{X}  \to  Z_B \to   \bar{q} q}$.}
\label{fig:1}
          \end{figure}
      \end{minipage}
      \hspace{0.05\linewidth}
      \begin{minipage}{0.46\linewidth}
          \begin{figure}[H]
              \includegraphics[width=\linewidth]{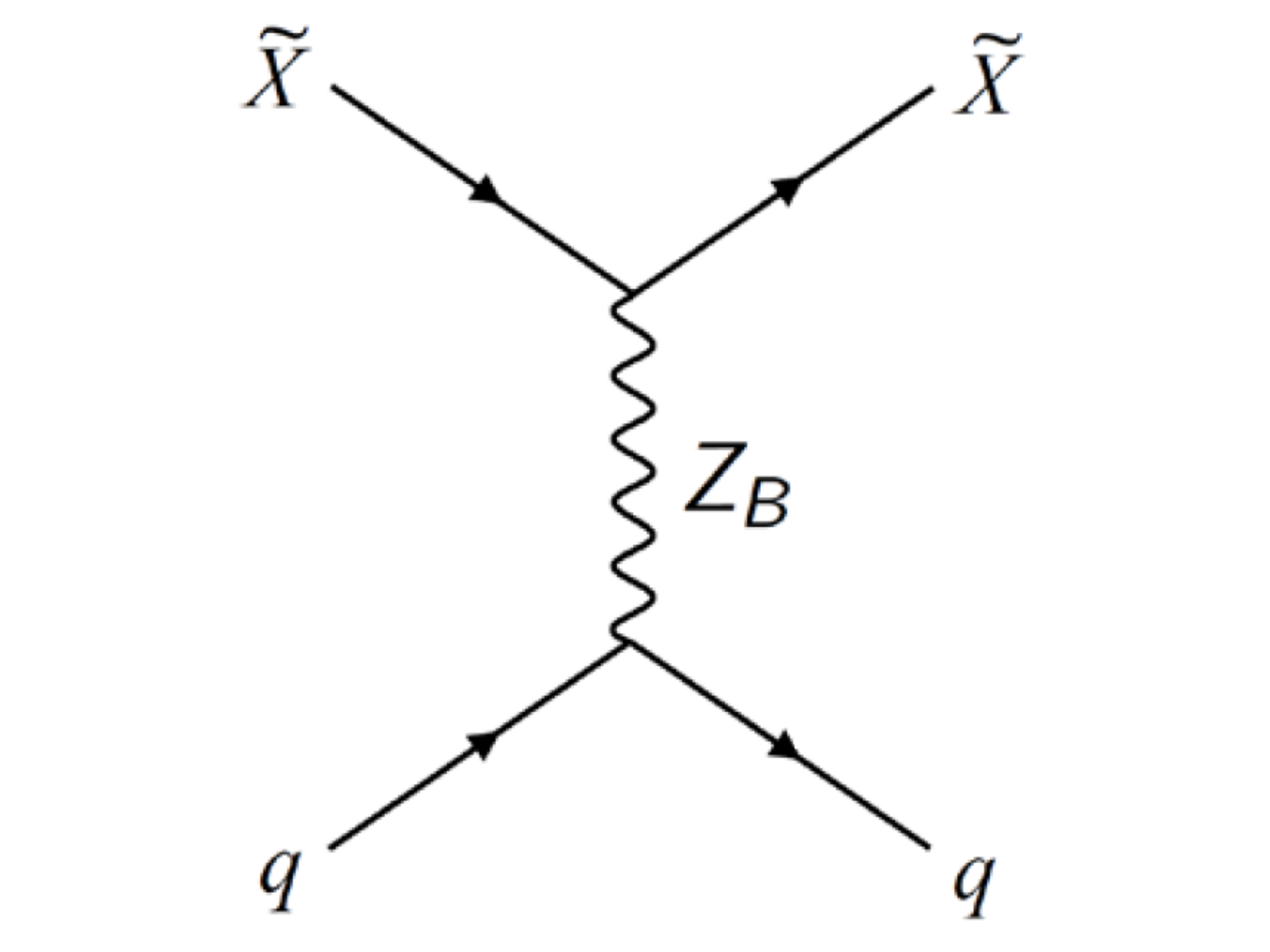}
\caption{Direct detection diagram $ {\tilde{X} N  \to  Z_B  \to \tilde{X} N}$. }
\label{fig:2}
          \end{figure}
      \end{minipage}
  \end{minipage}
  
  \vspace{1mm}

\begin{figure}[h!]
\centering
 \includegraphics[trim=0cm 1.3cm 0cm 0cm, clip=true, width=0.85\linewidth]{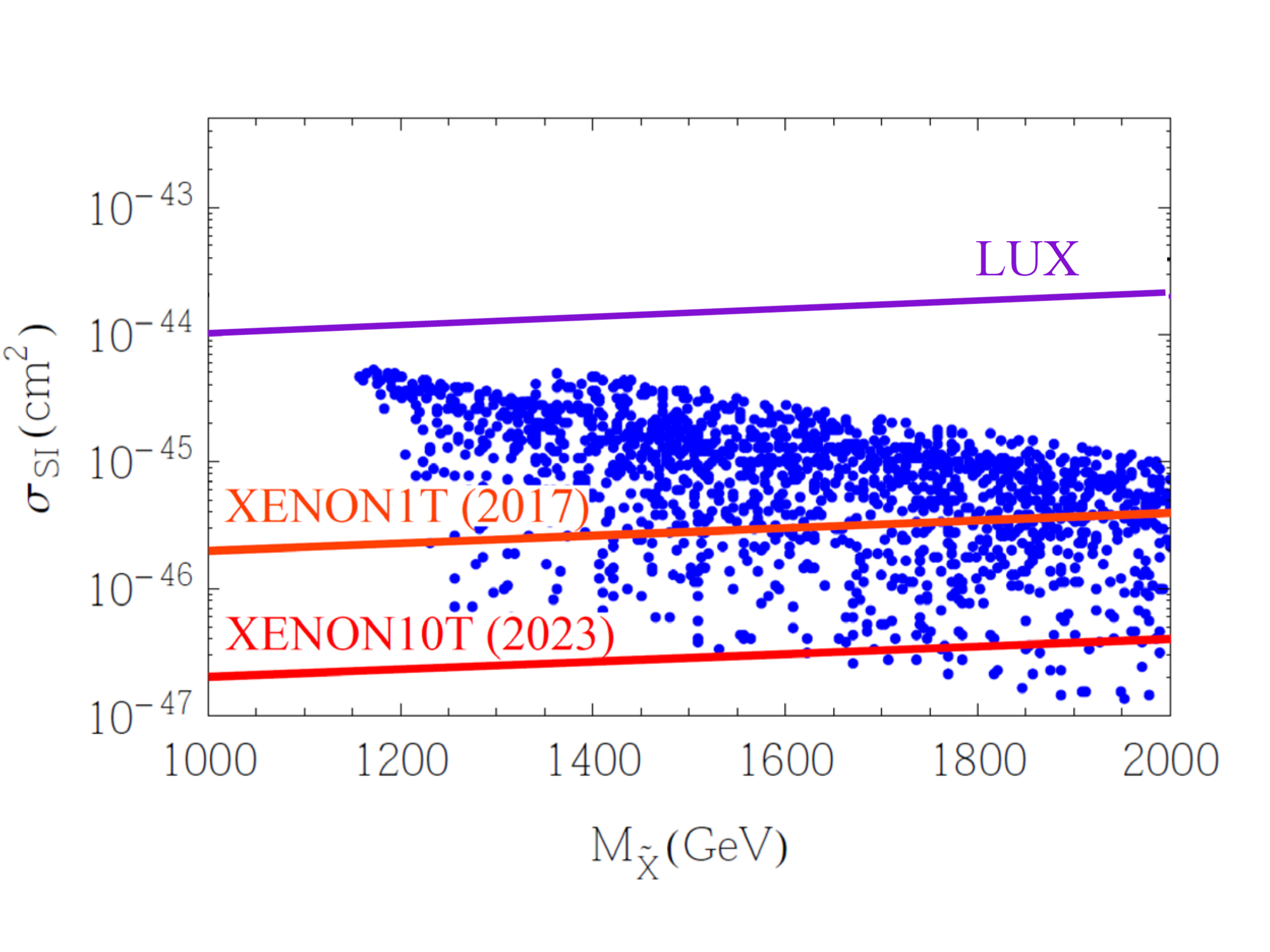}
 \caption{Direct detection cross section vs. DM mass for a scan over the parameters of the model (details are provided in the text) with overplotted current and future experimental bounds. \vspace{3mm} } 
 \label{fig:3}
 \end{figure}
 
 \vspace{10mm}
 
 Having calculated the two cross sections, we performed a scan over a range of parameter values of the model to see which points fulfill the experimental constraints.  A plot of such a scan showing the spin-independent direct detection cross section versus the DM mass is shown in Fig.~\ref{fig:3}. 
 For this particular parameter scan, we chose a concrete combination of couplings and quantum numbers equal to 1 (which is its most natural value), we varied the baryon number gauge coupling $0.1 \leq g_B \leq 0.3$, the $Z_B$ gauge boson mass $2.5 \ {\rm TeV} \leq M_{Z_B} \leq 5 \ {\rm TeV}$, and the relic density in the experimentally favored range $0.11 < \Omega_{\tilde{X}} h^2 < 0.13$. 
  The purple line denotes the latest DM bound coming from the LUX experiment and does not exclude any of the points. However, the XENON1T sensitivity projected for 2017 and the reach of XENON10T predicted for 2023 (red lines) will probe the majority of the scan points from the plot. Therefore, if there is no direct detection signal discovery within the next decade, the model will require a higher SUSY breaking scale in order to explain DM.

\section{Summary}

We have discussed an extension of the minimal supersymmetric standard model with gauged baryon and lepton numbers. This model has a number of attractive features. Most importantly, due to the extended gauge structure, it entirely forbids proton decay, even at the nonrenormalizable level. Baryon and lepton number symmetries are broken at the supersymmetry breaking scale. An experimentally viable scenario requires also violation of $R$-parity. This leads to many interesting lepton number violating signatures at colliders. Although the lightest supersymmetric particle  is no longer a good dark matter candidate, the model contains a new type of dark matter particle at the TeV scale carrying both baryon and lepton numbers. Its existence can be tested in dark matter experiments, as well as through missing energy signals at the LHC. We showed that the model accommodates a light active neutrino spectrum consistent with observation and predicts the existence two light sterile neutrinos. The measured values of mixing angles can be accommodated as well, however, the predictive power of the model still has to be strengthened in this regard.

\section*{Acknowledgements}
I am grateful to my collaborators: Pavel Fileviez P\'{e}rez, Sogee Spinner and Jonathan Arnold, without whom the work I discussed would not have been completed. 
I would also like to thank the organizers of the International Conference on Massive Neutrinos in Singapore, especially the chairman, Harald Fritzsch, for the invitation, warm hospitality, and a fantastic scientific and social atmosphere during the conference. This research was supported in part by NSF grant PHY-1316792.  

\bibliographystyle{ws-procs975x65}



\end{document}